\documentclass
[prb,aps,twocolumn,dcolumn,superscriptaddress,showpacs,amsfonts,amsmath,amssymb,showkeys,floatfix]{revtex4}
\usepackage{bm}
\usepackage{pstricks}
\usepackage{setspace}
\usepackage{graphicx}
\begin{document}
\title{Monte Carlo study of the spin-glass phase of the site-diluted dipolar Ising model}
\date{\today}
\author{Juan J. Alonso}
\affiliation{F\'{\i}sica Aplicada I, Universidad de M\'alaga,
29071-M\'alaga, Spain}
\email[E-mail address: ] {jjalonso@uma.es}
\author{Julio F. Fern\'andez}
\affiliation{Instituto de Ciencia de Materiales de Arag\'on, CSIC-Universidad de Zaragoza, 50009-Zaragoza, Spain}
\email[E-mail address: ] {jefe@Unizar.Es}

\date{\today}
\pacs{75.10.Nr, 75.10.Hk, 75.40.Cx, 75.50.Lk}

\begin{abstract}
By tempered Monte Carlo simulations, we study site-diluted Ising systems of magnetic 
dipoles.
All dipoles are randomly placed on a fraction $x$ of all $L^3$ sites of a simple cubic lattice, and
point along a given crystalline axis. 
For $x_c< x\le1$, where $x_c\simeq0.65$, we find an antiferromagnetic phase below a temperature
which vanishes as $x  \to x_c$ from above. At lower values of $x$, we find
an equilibrium spin-glass (SG) phase below a temperature given by 
$k_B T_{sg} \simeq x \varepsilon_d$, where $\varepsilon_d$ is a 
nearest neighbor dipole-dipole interaction energy. We study (a) the relative mean square deviation
$\Delta_q^2$ of $|q|$, where $q$ is the SG overlap parameter, and (b) $\xi_L/L$, where
$\xi_L$ is a correlation length. From their variation with temperature and system size, we determine $T_{sg}$.
In the SG phase, we find (i) the mean values $\langle \mid q\mid\rangle$ and $\langle q^2\rangle$ decrease
algebraically with $L$ as $L$ increases, (ii) double peaked, but wide, distributions of $q/\langle \mid q\mid\rangle $
appear to be independent of $L$, and (iii) $\xi_L/L$ rises with $L$ at constant $T$, but extrapolations to $1/L\to 0$
give finite values. All of this  is consistent with quasi-long-range order in the SG phase.
\end{abstract}

\maketitle

\section{Introduction}
\label{intro}
The collective behavior of spin systems in which magnetic dipole-dipole interactions
dominate has become the subject of considerable attention.  These systems are rare in nature, although some
ferroelectrics, \cite{ferroel}
and magnetic crystals such as LiHoF$_4$, an insulating magnetic salt,
have been known for decades to be well described by models of magnetic
dipoles. \cite{jensen,bel,grif} Much of the renewed interest in systems of
interacting dipoles comes from the experimental realization of
magnetic nanoparticle \cite{nanoscience} arrays \cite{np, sachan}  and of crystals of organometallic
molecules. \cite{nanomag} In these systems, particles up to some thousand Bohr magnetons
behave as single spins.  When closely packed
in crystalline arrangements,  dipolar interactions between them may
induce magnetic ordering. \cite{sachan,ordering}

Anisotropy also plays an important role in ordering dipolar systems.
The barrier energies $E_a$ that must be overcome by spins in order to reverse their direction
are often somewhat larger than the relevant dipolar energies $E_d$.
Then, collective effects can be observed when thermal energies are not sufficiently large to completely freeze
spins directions.
Their main effect is then to force spins to point up or down
along the easy  magnetization axis. \cite{rosen} Crystalline Ising dipolar systems (IDSs)
are then reasonable models. \cite{jensen}  
These systems are clearly frustrated,
since two different dipoles give rise to magnetic fields at any given site that are not in general
collinear. Not surprisingly, IDSs are very sensitive to their spatial arrangement.
Early work by Luttinger and Tisza  established which type of magnetic order arises at low
temperature for IDSs in each of the cubic lattices. \cite{lutt} More recently, we have obtained similar results by
much simpler methods. \cite{odip}  For instance, BCC and
LiHoF$_4$ like crystals are ferromagnetic ordered, but antiferromagnetic (AF) order
obtains on simple cubic (SC) lattices.  Competition between different
interactions brings about a more exotic magnetic order, known as ``spin ice",\cite{ice} in diamond crystals.

Whether disorder in IDSs, together with the
geometric frustration that comes with the dipolar interactions give rise to a thermodynamic spin-glass (SG) phase, is an interesting
question.\cite{vil}
Many experiments\cite{experiments} as well as  numerical simulations\cite{simul} have shown that assemblies of
classical magnetic moments placed at random,  such as in frozen ferrofluids and
diluted ferroelectric materials, exhibit  the time dependent behavior,
such as non-exponential relaxation and aging, \cite{memory}
that is expected from SGs.
However, search for evidence for the existence of an equilibrium SG phase
has been hampered by the extremely
slow relaxation that is inherent to these systems. In recent papers, we have given
numerical evidence that supports the existence of an equilibrium SG phase in IDSs with randomly
oriented axes both in fully occupied\cite{rad} and in partially occupied SC lattices.\cite{rad2}

Site dilution is a rather simple way to introduce disorder in 
experimental realizations of IDS. Some early attempts to find
a SG phase in Eu$_x$Sr$_{1-x}$S led to negative results.\cite{kotz}
By far the most scrutinized system for the last two decades has been LiHo$_x$Y$_{1-x}$F$_4$. In it,  
magnetic Ho$^{3+}$  ions are substituted, with little distortion, by 
non magnetic Y$^{3+}$ ions. \cite{bel} A strong uniaxial anisotropy
forces all spins to point up or down along the same axis at low temperatures.
This parallel-axis-dipolar (PAD) system orders ferromagnetically
a low temperature phase above
$x_c\simeq 0.25$. Below $x_c$, transitions from a paramagnetic to a SG
phase have been reported\cite{Rosen2, quilliam,ultimoq},
but the opposite conclusion, that no such transition takes place, has been reached in Ref. \onlinecite{barbara}. 
The issue is further obscured by quantum effects that may take place at $x\ll1$.\cite{gosh}

Theoretical results suggest that diluted PAD models undergo a SG
transition at low concentrations.  
An earlier study of bond-diluted Ising systems with 
long-range interactions (including the dipolar case) found that SG order may exist 
at low temperatures in the limit of weak concentration. \cite{aha} 
Mean field calculations for  site-diluted PAD systems in FCC and BCC lattices 
predicted a SG phase for concentrations $0<x<x_c$ where $x_c$ is the value above which
ferromagnetic order ensues. \cite{xu} 
More recently, Edwards-Anderson \cite{ea} type models with 
power-law decaying interactions $J_{ij}\sim1/r_{ij}^\sigma$ have been 
studied. \cite{bray,  katz}
A $1$D Ising Spin Glass model has been found to have a nonzero temperature
SG phase transition for $\sigma<1$.\cite{katz}  A $3$D Ising systems 
with RKKY interactions (that decay with $1/r_{ij}^3$) 
have been predicted  to lie in the same universality class 
as the $3$D Ising Edwards-Anderson (EA) model with short range 
interactions. \cite{bray} 

Numerical methods have provided conflicting answers
to the question of the existence of a SG phase in site diluted PAD models. 
Biltmo and Henelius\cite{bh} have calculated that the ferromagnetic phase of  LiHo$_x$Y$_{1-x}$F$_4$
extends down to $x_c\simeq 0.24$, but found no SG phase at low temperatures for $x<x_c$.\cite{bh} This is 
 in contradiction with another MC simulation for the same system that finds a SG transition 
 for concentrations $x=0.065$ and  $0.125$. \cite{gin} 
 Numerical work has also been done on a PAD model 
on a SC lattice, using a Wang-Landau 
MC method.\cite{yu} No transition was found for  $x\le0.2$. 

Here we also simulate
a PAD model on a SC lattice.
Our justification for working with a SC lattice is as follows.
Whereas such systems order AF in fully occupied SC lattices,\cite{lutt,odip} instead of ferromagnetically, as in the LiHoY$_4$ lattice, 
the physics of PAD systems is not expected to depend on lattice structure for $x\ll 1$. A continuum should then lead
to the same behavior. Furthermore, rescaling distance $r$ as $r \to r/\rho^{1/3}$, where $\rho$ is
the spatial density of spins, 
is no different from redefining dipolar energies by $\varepsilon_d\to
\rho \varepsilon_d$, since dipolar interactions decay as $r^{-3}$. Now,
consider $k_BT_{sg}/n_d\varepsilon_d$ for any lattice structure, where $k_B$
is Boltzmann's constant, $T_{sg}$ is the SG transition temperature, $n_d$ is the
number of magnetic dipoles within a $d^3$ volume, and $\varepsilon_d$ is the 
smallest possible dipolar energy two parallel dipoles that are a distance $d$ apart can have.
Clearly, $k_BT_{sg}/n_d\varepsilon_d $ must be independent of lattice structure for $x\ll1$.
This enables us to compare results for SC and LiHoF$_4$ lattices, or any other lattice,
for $x\ll 1$. Such a comparison is made in Table I.
 
\begin{table}\footnotesize
\caption{Spin-glass transition temperature for PAD systems. NIL is entered where a transition has been concluded not to take place.
For LiHo$_{x}$Y$_{1-x}$F$_4$, we let $d = 5.175$ \AA, hence the mean number of spins in volume $d^3$ is $n_d=1.926x$ (since unit cells of LiHoYF$_4$ are
$ 5.175\times 5.175\times 10.75$ \AA$^3$ large and have 4 Ho ions each\cite{bel}); furthermore, $\varepsilon_d=0.214$ K.\cite{ed} On simple cubic lattices, we let $d=a$, hence $n_d=x$. $\chi_3$ is the nonlinear susceptibility,  and $\nu$ is the critical exponent for the correlation length.  }
\begin{ruledtabular}
\begin{tabular}{c}
On LiHoYF$_4$ type lattices   \\
\end{tabular}
\begin{tabular}{|c|c|l|c|c|c|}
Ref.  & Method & $x$  & $n_d\varepsilon_d $   &$k_BT_{sg}/n_d\varepsilon_d $ & $\nu$  \\
\hline
\onlinecite{Rosen2} & $\chi_3$ & $ 0.167$ & 0.069 K &  1.9 &\\
\onlinecite{quilliam} & $\chi_3$    & $0.045$ & $0.019$ K& 2.3 & \\
\onlinecite{ultimoq} & $\chi_3$  & $0.167$ & 0.069 K& 3.1  &\\
\onlinecite{barbara} &  $\chi_3 $ & $ 0.165$ &  0.068 K& NIL & \\
\onlinecite{barbara} &  $\chi_3 $ & $ 0.045$ & 0.019 K& NIL & \\
\onlinecite{bh}  & MC  & $0.06 $ & 0.025 K & NIL & \\
\onlinecite{bh}  & MC  & $ 0.12 $ & 0.049 K & NIL & \\
\onlinecite{gin} &MC & $ 0.125$ & 0.052 K  & 1.8 & $1.3$  \\
\onlinecite{gin} &MC & $ 0.0625$ &0.026 K & 1.6  &  $1.3$ \\
\end{tabular}
\begin{tabular}{c}
On simple cubic lattices \\
\end{tabular}
\begin{tabular}{|c|c|l|c|c|}
Ref.  & Method & $x$     &$k_BT_{sg}/n_d\varepsilon_d $  & $\nu$ \\
\hline
\onlinecite{yu} & MC & $0.045, 0.12, 0.20$ &  NIL & \\
here & MC & $0.35$ &   1.0(1) & $0.95$ \\
here & MC & $0.20$ &   1.0(1) & $0.95$\\
\end{tabular}
\end{ruledtabular}
\end{table}

The main aim of this paper is to find, by means of MC simulations, whether an 
equilibrium  SG phase exists in site diluted systems of dipoles, which are
placed at random on the sites of a SC lattice and
point up or down along a chosen principal axis.
Since in the limit of low 
concentrations details of the lattice are expected to become
irrelevant, our results have direct connection  with the experimental and numerical
work mentioned above. In this regard, we follow along the lines of Ref. \onlinecite{gin}.
But we aim to go further. It is our purpose to also find whether  the SG phase of the PAD model behaves marginally, that is,
it has quasi-long-range order (as the $XY$ model \cite{xy} in 2D), or whether it has spatially-extended states,\cite{sinova} as in the droplet\cite{droplet}
and replica-symmetry-breaking\cite{RSB} pictures of the SG phase.

The plan of the paper is as follows. In Sec. \ref{mm} we define the model, 
give details  on how we apply the parallel tempered Monte Carlo 
(TMC) algorithm, \cite{tempered} in order to get equilibrium results.
We also define the quantities we calculate, including the spin overlap\cite{ea} $q$, and $\xi_L$, often referred to as a ``correlation length''.\cite{longi,balle,katz0}
In Sec. \ref{resultsA} we give results for the dipolar AF phase we obtain for $x>x_c$, where $x_c\simeq 0.65$, as well as for
its nature and boundary.
In Sec. \ref{resultsB}, we give numerical results we have found for (i) $q$ distributions and (ii) $\xi_L/L$, within the following $x$ and $T$ ranges, 
$0.2\leq x < 0.65$ and $0.6x \lesssim T \leq 1.5x$. In Sec. \ref{valueTsg} we examine the evidence we have in favor of the existence of
a paramagnetic to SG phase transition when $x<x_c$, and find that the transition temperature is given by
$k_BT_{sg} \simeq x\varepsilon_d $, where $\varepsilon_d$ 
is a nearest neighbor dipole-dipole interaction energy which is defined in Sec. \ref{mm}.
In order to study the nature of the SG phase, we examine the following evidence in Sec. \ref{wlro}:
(i) the mean values $\langle \mid q\mid\rangle$ and $\langle q^2\rangle$ decrease
algebraically with $L$ as $L$ increases, (ii) double peaked, but wide, distributions of $q/\langle \mid q\mid\rangle $
appear to be independent of $L$, and (iii) $\xi_L/L$ rises with $L$ at constant $T$, but extrapolates to finite values as $1/L\to 0$.
We provide a specific example of spatial correlation functions which decay algebraically with distance
but lead to $\xi_L/L$ curves that spread out with $L$ (for finite values of $L$) as $T$ decreases below $T_{sg}$,
in rough agreement with our MC results for $\xi_L/L$.  All of this is consistent with
quasi-long-range order in the SG phase. In Sec. \ref{nuexp} we find the best pair of values for $T_{sg}$ and $\nu$, to have curves $\xi_L/L$
for various values of $L$ collapse onto a single curve if plotted vs $(T/T_{sg}-1)L^{1/\nu}$ over the $T>T_{sg}$ range. The values
given in Table I are obtained.

\section{model, method, and measured quantities}
\label{mm}
\subsection{Model}
We consider site-diluted systems of Ising magnetic dipoles  on a SC lattice. 
All dipoles point along the $z$ axis of the lattice. Each site is occupied with probability $x$. 
The Hamiltonian is given by,
\begin{equation}
{\cal H}=\frac{1}{2}\sum_{ ij}
T_{ij}\sigma_i \sigma_j
\end{equation}
where the sum is over all occupied sites $i$ and $j$ except $i=j$, 
$\sigma_i=\pm 1$ on any 
occupied site $i$, 
\begin{equation}
T_{ij}=\varepsilon_a
(a/r_{ij})^3(1-3
z_{ij}^2/r_{ij}^2),
\label{T}
\end{equation} 
$ r_{ij}$ is the distance between $i$ and $j$ sites, $z_{ij}$ is the $z$ component of $ r_{ij}$, $\varepsilon_a$ is an energy, and $a$ is the
SC lattice constant. In the following  we give all  temperatures and energies in terms of $\varepsilon_a/k_B$ and $\varepsilon_a$, respectively.
Hence, $k_BT/n_a\varepsilon_a=T/x$ from here on. 

This model is  clearly an Ising model with long-range interactions where
bond strengths $T_{ij}$ are determined by the  dipole-dipole terms. Note that 
$T_{ij}$ signs are not distributed at random, but depend only on the
orientation of vectors ${\bf r}_{ij}$ on a SC lattice. 
This is to be contrasted with a  {\it random-axes} dipolar model, (RAD)\cite{rad}  in 
which Ising dipoles point along directions ${\bf n}_i=(n^\alpha_i, \alpha=1,2,3)$  that are chosen  
at random by sorting two independent random numbers for each site, introducing randomness on bond
strengths $T^{\alpha \beta}_{ij}$. This is why PADs exhibit  AF order at high concentration in
contrast with RADs, that do not. \cite{rad}

\subsection{Method}

We use periodic boundary conditions (PBC). As is usual for PBC, think of a periodic arrangement of replicas that span all space beyond the system of interest.
These replicas are exact copies of the Hamiltonian and of the spin configuration of the system of interest.  
Details of the PBC scheme we use can be found in Ref. \onlinecite{odip}.
We let a spin on site $i$ interact through dipolar fields with all spins within an $L\times L\times L$
cube centered on it. No interactions with other spins are taken into account.
This introduces an error which we show in Appendix I to vanish as $L\rightarrow \infty$, regardless
of whether the system is in the paramagnetic, AF or SG phase. There is,
therefore, no effect on the thermodynamic limit of the system of interest here. (The result
we obtain in Appendix I is not applicable to an inhomogeneous ferromagnetic- phase or critical region- that may obtain on other lattices.)

\begin{table}\footnotesize
\caption{Parameters of the tempered MC simulations. $x$ is the probability that any given site is occupied by a magnetic dipole; $L$ is the linear lattice size; $\Delta T$ is the temperature step in the TMC runs; $T_o$ and $T_n$ are the highest and the lowest temperatures, respectively; $N_r$ is the number  of (quenched) disordered samples; a number $t_0$ of MC sweeps are made before any measurements are taken. The measuring time interval is $[t_0, 2t_0]$ in every case.}
\begin{tabular}{p{0.8cm} p{1.0cm} p{1.0cm} p{1.0cm} p{1.5cm} p{1.5cm}}
\hline\hline
\multicolumn{6}{c}
{$x=0.20$, $\Delta T=0.02$, $T_0=0.8 $ }\\
\hline
  L& 4   &  6  &8 & 10 \\
   $T_n$ & 0.06 &0.06& 0.06&0.12\\
   $N_r$ & 8500 & 3800 & 1000& 800\\
   $t_0$ & $5 \times  10^7$ & $5 \times  10^7$& $5 \times  10^7$& $5 \times  10^7$\\
 \hline 
\multicolumn{6}{c}
{$x=0.35$, $\Delta T=0.05$, $T_0=2.0 $ }\\
\hline
  L& 4   &  6  &8 & 10 &12\\
   $T_n$ & 0.05 &0.05& 0.05&0.275 & 0.35\\
   $N_r$ & 9000 &5000 & 1100& 380 & 200\\
   $t_0$ & $4 \times  10^6$ & $4 \times  10^6$ & $ 4 \times 10^6$& $4 \times  10^6$&$4 \times  10^6$\\ 
\hline
\multicolumn{6}{c}
{$x=0.50$, $\Delta T=0.05$, $T_0=2.0 $}\\
\hline 
  L& 4   &  6  &8 & 10 \\
   $T_n$ & 0.1 &0.05& 0.05& 0.35\\
   $N_r$ & 1000 &650 & 500& 300 \\
   $t_0$ & $5 \times  10^5$ & $5 \times  10^5$ & $ 4\times10^6$& $  10^7$\\
\hline 
\multicolumn{6}{c}
{$x=0.60$, $\Delta T=0.1$, $T_0=2.0 $}\\
\hline 
  L& 4   &  6  &8 & 10 \\
   $T_n$ & 0.10 &0.10& 0.20& 0.30\\
   $N_r$ & 1400 &500 & 800& 300 \\
   $t_0$ & $4 \times  10^6$ & $4 \times  10^6$ & $ 4\times10^6$& $ 4\times10^6$\\
\hline 
\multicolumn{6}{c}
{$x=0.65$, $\Delta T=0.1$, $T_0=3.0 $}\\
\hline 
  L& 4   &  6  &8 & 10 \\
   $T_n$ & 0.10 &0.10& 0.10& 0.30\\
   $N_r$ & 1400 &900 & 1400& 540 \\
   $t_0$ & $4 \times  10^6$ & $4 \times  10^6$ & $ 4\times10^6$& $ 4\times10^6$\\
\hline 
\multicolumn{6}{c}
{$x=0.70$, $\Delta T=0.1$, $T_0=3.0 $}\\
\hline 
  L& 4   &  6  &8 & 10 \\
   $T_n$ & 0.10 &0.10& 0.10& 0.30\\
   $N_r$ & 750 &200 & 100& 100 \\
   $t_0$ & $4 \times  10^6$ & $4 \times  10^6$ & $ 4\times10^6$& $10^6$\\
\hline
\multicolumn{6}{c}
{$x=0.75$, $\Delta T=0.1$, $T_0=3.0 $}\\
\hline
  L& 4   &  6  &8 & 10 \\
   $T_n$ & 0.10 &0.10& 0.10& 0.10\\
   $N_r$ & 1000 &200 & 100& 100 \\
   $t_0$ & $4 \times  10^6$ & $4 \times  10^6$ & $ 2 \times10^6$ & $10^6$\\
\hline 
\multicolumn{6}{c}
{$x=0.80$, $\Delta T=0.1$, $T_0=3.0 $}\\
\hline 
  L& 4   &  6  &8 & 10 \\
   $T_n$ & 0.10 &0.10& 0.10& 0.10\\
   $N_r$ & 600 &200 & 220& 100 \\
   $t_0$ & $4 \times  10^6$ & $4 \times  10^6$ & $ 10^6$ & $10^6$\\
\hline\hline 
\end{tabular}
\label{table}
\end{table}

In order to bypass energy barriers that can trap a system's state at low temperatures in the glassy phase 
we have used  the parallel tempered Monte Carlo (TMC) algorithm.\cite{tempered,ugr}
We apply the TMC algorithm as follows. We run in parallel a set of $n$ identical systems at 
equally spaced temperatures $T_i$, given by $T_i=T_0- i \Delta T$ where $i=0, \cdots, n-1$
and $\Delta T>0$.  By {\it identical} we mean here that all $n$  systems have the same quenched distribution
of empty sites, though each system starts from an independently chosen initial condition. We apply the TMC algorithm 
to any given system in two steps. In the first step, system $i$ evolves independently for 8 MC sweeps
under the standard single-spin-flip Metropolis algorithm.\cite{mc}  (Owing to dipolar interactions,  the
MC sweep time scales as $N^2$, where $N$ is the number of spins.) We update all dipolar fields throughout the system every time a  spin flip is accepted. In the second step, we give system $i$ a chance to exchange states with 
system $i+1$ evolving  at a lower temperature $T_i-\Delta T$. We accept exchanges with probability 
$P=1$ if $\delta E= E_{i}-E_{i+1}<0$, and $P=\exp (-\Delta \beta \delta E)$ otherwise, where 
$\Delta \beta=1/T_{i+1}-1/T_{i}$. The cycle is complete when
 $i$ has been swept from $0$ to $n-2$.  Thus, we associate eight MC sweeps with each cycle.
For the simulation to converge at low temperatures it is
important to choose $\Delta T$ small enough to allow frequent state exchanges between systems.  
This will often be fulfilled if $\Delta \beta\Delta E\lesssim 1$. 
The required condition, $\Delta T \lesssim T/\sqrt{Nc}$, follows for $\Delta T$ 
where $c$ is the specific heat per spin.  Then, we obtain appropriate
values for  $\Delta T$ from inspection of plots  of the specific heat vs $T$. \cite{rad} 
We find it helpful to have the 
highest temperature $T_0$ at least twice as large as 
what we expect to be the transition temperature between the paramagnetic and 
the ordered phase for obtaining equilibrium results in the ordered phase.

In our simulations the  $n$ identical systems 
start from completely disordered spins configurations.
We need equilibration times $t_0$ of at least $4 \times 10^6$ MC sweeps 
for $x\le0.7$ for systems with a number dipoles $N\ge200$ (see at  the end of this sections for details on 
how we choose $t_0$). Thermal averages come from averaging over the time range $[t_0,2 t_0]$.  We further
average over $N_r$  samples
with different realizations of disorder. 
Values of the parameters for all TMC runs are given in Table I.

\subsection{ Measured quantities}
\label{meas}
We next specify the quantities we calculate.  
We obtain the specific heat from the temperature derivative of the energy.
For the staggered magnetization, we define, as  befits a PAD model  on a SC lattice\cite{odip}
\begin{equation} 
m=N^{-1} \sum_i \sigma_i (-1)^{x(i)+y(i)}
\label{phi}
\end{equation}
where $x(i)$ and $y(i)$ are the space coordinates of site $i$. We calculate the probability
distribution $P_m$, as well as the moments 
\begin{equation}
m_n\equiv \langle |m|^n\rangle ,
\label{moments}
\end{equation}
for $n=1,2$, where $\langle ...\rangle$ stand for averages over time and over a number $N_r$ of system samples with different quenched disorder. 
Unless otherwise stated, time averages are performed over a time range $t_0<t<2t_0$, and $t_0$ is chosen as specified below in order to ensure equilibrium. 
We make use of these moments to calculate the staggered susceptibility and the mean square deviation of $\mid m\mid /m_1$, that is,
\begin{equation}
\Delta^2_m=\frac{m_2}{m_1^2}-1.
\label{deltam2}
\end{equation}
In order to spot SG behavior, we also calculate
the Edwards-Anderson overlap parameter,\cite{ea}
\begin{equation} 
q=N^{-1} \sum_j \phi_j
\label{q}
\end{equation}
where 
\begin{equation} 
\phi_j=\sigma^{(1)}_j\sigma^{(2)}_j
\label{phil}
\end{equation}
$\sigma^{(1)}_j$ and $\sigma^{(2)}_j$ are the spins on site $j$ of identical replicas $(1)$ and $(2)$ of the system of interest. As usual, identical replicas have the same Hamiltonian, and are at the same temperature, but are in uncorrelated states.
Clearly,  $q$  is a measure of the spin configuration overlap between the two replicas. As we do for $m$, we calculate the probability distribution $P_q$ as well as the moments  
$q_1=\langle |q|\rangle $ and $q_2=\langle q^2\rangle $, in analogy to Eq. (\ref{moments}). The SG susceptibility $\chi_{sg}$ is given by $Nq_2$. Finally, we also make use of the relative mean square deviation of $q$, $\Delta^2_q=q_2/q_1^2-1$.

We need to make sure that equilibrium is reached before we start taking measurements. 
To this end, we define a time dependent spin overlap $\tilde {q}$, not between pairs of 
identical systems, but between spin  configurations of the same system at two different 
times $t_0$ and $t_1=t_0+ t$ of the same TMC run,
\begin{equation} 
\tilde q(t_0,t)=N^{-1} \sum_j \sigma_j(t_0)\sigma_j( t_0+t).
\label{phi0}
\end{equation}

\begin{figure}[!t]
\includegraphics*[width=80mm]{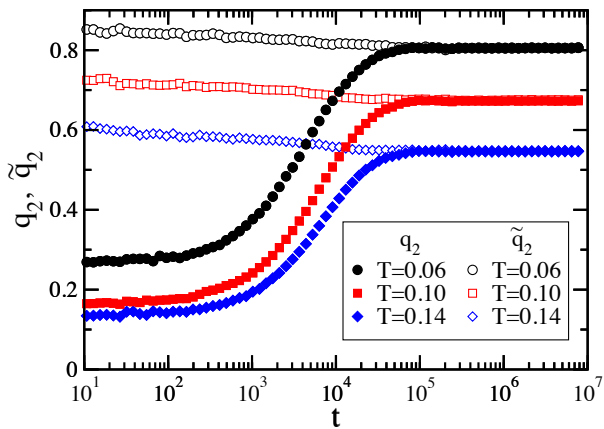}
\caption{(Color online) 
Semilog plots of $\tilde q_2(t_0,t)$ and $q_2$ vs time $t$ (in MC sweeps) for systems of $8 \times 8 \times8$ spins 
at the values of $T$ shown in the figure. Here, $q_2$ comes from averages of $ q^2$ over time,  starting at $t=0$ from an initial random spin configuration.
Here, $t_0=10^7$ MC sweeps. A data point at time $t$ stands for an average over a time interval $[t,1.2t]$  and
over $10^3$ system samples.
}
\label{figcero}
\end{figure}

Let $\tilde{q}_2(t_0,t)=\langle  [  \tilde q(t_0,t)  ]^2 \rangle$. Suppose thermal equilibrium is reached long before time $t_0$ has elapsed. 
Then, $\tilde{q}_2(t_0,t)\rightarrow q_2$ at some time $t$ long before $t=t_0$. Plots of $\tilde{q}_2(t_0,t)$ vs $t$, for $10^{-6}t_0<t<t_0$, for $t_0=10^7$ MC sweeps, are shown in Fig.~\ref{figcero} for $x=0.20$ and various values of $T$.
Plots of $q_2$, obtained by averaging $ q^2$ over time, not starting at $t=t_0$, as we do everywhere else in order to obtain equilibrium values, but starting at $t=0$, from an initial random spin configuration, 
are also shown in Fig.~\ref{figcero} for comparison.  Note  that both quantities do become approximately equal when $t\gtrsim 10^5$ MC sweeps. In order to obtain equilibrium results, we have always chosen sufficiently large values  of $t_0$ to make sure that  $\tilde{q}_2(t_0,t)\rightarrow q_2$ long before $t=t_0$. All values of $t_0$ and $N_r$ are given in Table II.

As has become customary in SG work,\cite{longi,balle,katz0} we calculate quantity $\xi_L$,
\begin{equation} 
\xi^2_L=\frac {1 } {4 \sin^2  ( k /2)}  { \left[ \frac{\langle q^2 \rangle} { \langle\mid q({\bf k})  \mid ^2   \rangle}  -1 \right] }, 
\label{phi1}
\end{equation}
where
\begin{equation} 
q({\bf k})=N^{-1} \sum_j \phi_j e^{i {\bf k\cdot r}_j},
\label{phi2}
\end{equation}
${\bf r}_j$ is the position of site $j$, and ${\bf k} =(2\pi/L,0,0)$. Recall this system is anisotropic, interactions
along the spin axes are twice as large as in a perpendicular direction.
We have found this direction of  ${\bf k}$ (perpendicular
to all spin directions) to be more convenient to work with than the direction along the spin axes.

Note that replacement of $\exp ({i {\bf k.r}_j})$ by $ 1-i{\bf k.r}_j$ gives
\begin{equation}
\xi^2_L=  \frac { \sum_{ij}    [ \textbf {k}  \cdot  ( \textbf {r}_i- \textbf { r}_j)]^2 \langle \phi_i\phi_j\rangle }{8 \sin^2  ( k /2)\sum_{ij} \langle \phi_i\phi_j\rangle }.
\end{equation}
This is right in the $\xi_L/L\to 0$ limit. 
The above equation clearly shows that $\xi_L$ is then (up to a multiplicative constant) the spatial correlation length (in the \textbf{k} direction) of $\langle \phi_0\phi_r\rangle$. Therefore, we can think of $\xi_\infty$, the $L\to\infty$ limit of $\xi_L$, as the correlation length of a \emph{macroscopic} system in the \emph{paramagnetic} phase.
In a condensed phase, on the other hand, condensate fluctuations generally take place over finite lengths $\bar{\xi}$, but
$\xi_L/L\to \infty$ as $L\to \infty$ if there is \emph{strong} long-range  order, that is, if $\langle \phi_0\phi_r\rangle$ does not vanish as $r\to \infty$.
One would have to replace $\phi$ by $\phi -\langle \phi\rangle$  in Eq. (\ref{phi1}) in order to relate $\xi_\infty$ to $\bar{\xi}$. 
Following current usage, we shall nevertheless refer to $\xi_L$ as ``the correlation length''.

In contrast with $P_q$ and its first moments, $\xi_{L}$ takes into account spatial variations of the
EA overlap $q$ and is yet another probe for detecting a SG transition. \cite{longi,balle,katz0}

\section{The AF phase}
\label{resultsA}
\begin{figure}
\includegraphics*[width=80mm]{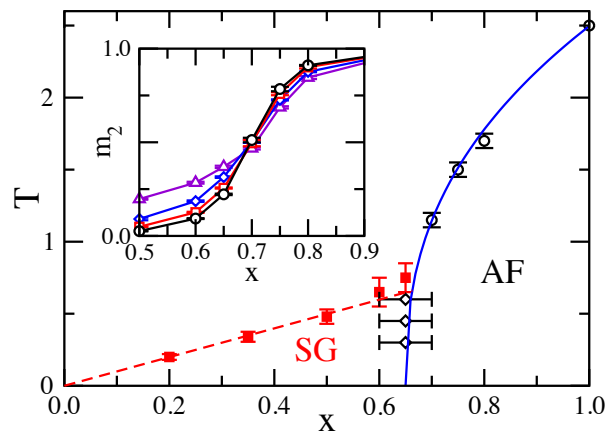}
\caption{(Color online) 
Phase diagram of the PAD model.
$\circ$ stand for the N\'eel temperature $T_{AF}$, and 
$\blacksquare$ stand for the SG transition temperature $T_{sg}$. 
$\diamond$ stand for maxima value of $x$ for which $m_2$ decreases as $N$ increases
for each of three fixed values of $T$. The full line for the phase boundary between the paramagnetic and 
AF phases is a fit to the data points, given by,  
$T_{AF} \simeq 3.8 (x-x_c)^{0.4}$, where $x_c=0.65$. 
 The straight dashed-line is for $T_{sg}=x\varepsilon_a$.
In the inset, $m_2$ versus $x$ for $T=0.4$.  
$\circ$, $\square$, $\diamond$, and $\triangle$,  stand for $L=10, 8, 6$, and $4$
 respectively. }
\label{1fases}
\end{figure}

Our main results for the PAD model are summarized in the phase diagram 
exhibited in Fig. \ref{1fases}. A thermally 
driven second order transition takes place at the phase boundary between the paramagnetic
and AF phases. The phase boundary meets the $T=0$ line at $x\simeq 0.65$. We shall
refer to the value of $x$ at this point as $x_c$.

In this 
section we report the numerical evidence for the paramagnetic-AF transition.\cite{DISx} Results having to do
with the spin glass are given in the next section.

\begin{figure}[!t]
\includegraphics*[width=85mm]{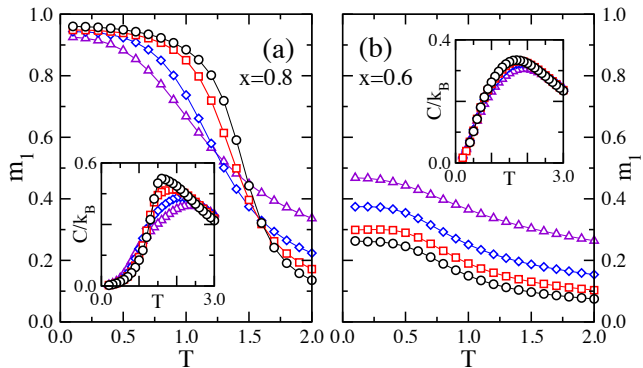}
\caption{(Color online) 
(a) Staggered magnetization $m_1$ vs $T$ for $x=0.8$. Icons
$\circ$, $\square$, $\diamond$, and   $\triangle$ 
stand for
$L=10, 8, 6$ and $4$ respectively. Lines are only guides to the eye.  Note $m_1$ grows with $L$
at low temperature, consistently with an AF phase. 
In the inset, specific heat vs. $T$ for the same values of $x$ and of system
sizes. The sharp variation $C$ with respect to $T$ near $T=1.5$ is consistent with an AF phase transition thereon. (b) Same as in (a) but for $x=0.6$. Note (i) $m_1$ decreases with $L$
at all temperatures, consistently with the nonexistence of an AF phase, and (ii) a rounded specific heat, consistent with a SG transition. In all panels, error bars are smaller
than symbol sizes. }
\label{2prelimi}
\end{figure}

\begin{figure}[!t]
\includegraphics*[width=85mm]{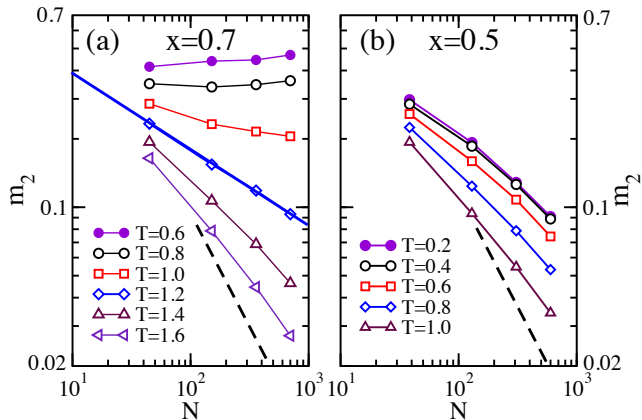}
\caption{(Color online) (a) Log-log plots of $m_2$ versus 
$N $ for $x=0.7$ and the values of $T$ shown. 
Continuous lines are guides to the eye, except for the straight line over the data points for $T=1.2$, which 
is for $1/N^{0.35}$. A dashed line shows the slope one expects for a macroscopic paramagnet.
(b) Same as in (a) but for $x=0.5$.  In all panels, error bars are smaller
than symbol sizes.  }
\label{3magne}
\end{figure}

The AF phase is defined
by the staggered magnetization, as given in Eq. (\ref{phi}). We illustrate in Fig. \ref{2prelimi}a how 
the staggered magnetization $m_1$ behaves with temperature for
$x=0.8$. This is in sharp contrast to the behavior of $m_1$
for small $x$, where an AF phase does not exist. 
Such behavior
is exhibited in Fig. \ref{2prelimi}b.
Note that  $m_1$ appears to decrease as $N$ increases even at low $T$. 
We obtain similar results for the staggered magnetization for other values of
$x$ (shown in Fig. \ref{1fases}) below $x_c$. 
This is our first piece of evidence for the nonexistence of an AF phase
below some $x_c$ and that $x_c\sim 0.6$. We return to this point in the discussion of Fig. \ref{3magne}.

\begin{figure}[!t]
\includegraphics*[width=85mm]{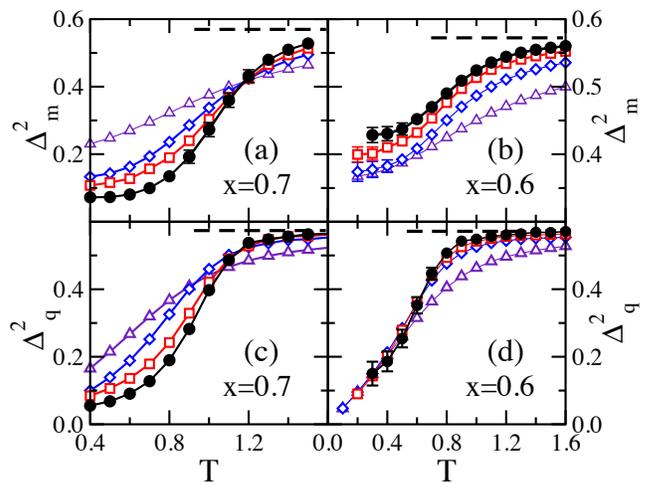}
\caption{(Color online) (a) Plots of $\Delta ^2_m$ vs $T$, for $x=0.7$. $\bullet$, $\square$, $\diamond$, and $\times$ are for $L=10, 8, 6$ and $4$, respectively. Lines are guides to the eye.  
The thick dashed-line is for the macroscopic paramagnetic limit  $\pi/2-1$. (b) Same as in (a) but for $x=0.6$. 
(c) Plots of  $\Delta ^2_q$ vs $T$, for $x=0.7$. Symbols are as in (a). 
(d) Same as in (c) but for $x=0.6$. Error bars are shown only where they are larger than symbol sizes.}
\label{4d2xalto}
\end{figure}

Plots of the specific heat $C$ vs $T$ are shown in the insets of Figs. \ref{2prelimi}a 
and \ref{2prelimi}b.
Note 
the sharp variation of $C$ vs $T$ near $T=1.5$, in Fig. \ref{2prelimi}a, as one expects from a paramagnetic-AF phase transition. 
Note also how, as one expects for a paramagnetic-SG transition, $C$ varies smoothly for a smaller value of $x$, in Fig. \ref{2prelimi}b.

For further information about the extent of the AF phase, we now examine how $m$ varies with $N$
for some values of $x$ and of $T$. 
Compare the log-log plots of $m_2$ versus the number of
dipoles $N$ on Figs. \ref{3magne}a and  \ref{3magne}b, respectively.
The data points in Fig. \ref{3magne}a are consistent with a
second order phase transition from a magnetically disordered phase, above
$T= 1.2(1)$, for which $N m_2 = O(1)$, to a strong long-range order below $T = 1.2(1)$,
where $m_2=O(1)$. Note  that $m_2 \sim 1/N^{p}$ at  $T=1.2$. From the definition of $\eta$ (see Sec. \ref{wlro} or
Ref. \onlinecite{mefisher}), $3p=1+\eta$ follows, which gives $\eta=0.05$. We are however not too interested here
in such details of the critical behavior on the $T=T_{AF}(x)$ line.
In Fig. \ref{3magne}b, $m_2$ vs $N$ plots show faster than algebraic decay with $N$.
This shows we are then beyond the bounds of the AF phase. We have followed this criterion
as a first approach in establishing the boundary of the AF phase. Plots of 
$m_1$ (instead of $m_2$) vs $N$ show the same qualitative behavior.

\begin{figure}[!b]
\includegraphics*[width=75mm]{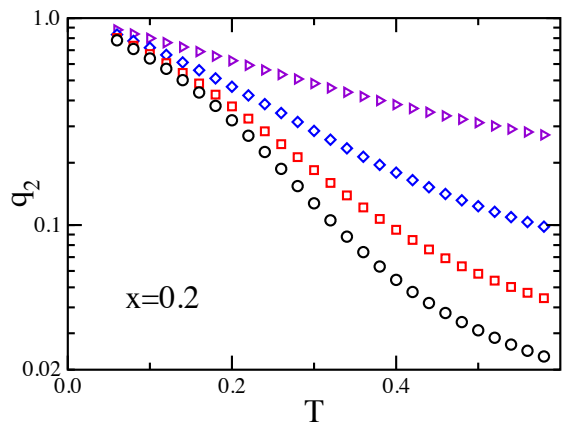}
\caption{ (Color online)
Semilog plots of $q_2$ versus $T$  for $x=0.2$,  and 
$L=10$, ($\circ$), $L=8$ ($\square$), $L=6$ ($\diamond$), and $L=4$ ($\triangleright$). 
All error bars are smaller than symbol sizes.  }
\label{5q2vsT}
\end{figure}

We draw more quantitative results about the AF phase boundary  from the behavior of the relative uncertainty $\Delta_m^2$. 
We first outline how we expect $\Delta_m^2$ to behave as a function of $T$ and $x$ in the various magnetic phases.
It clearly follows from its definition in Eq. (\ref{deltam2}) that $\Delta_m^2\rightarrow 0$ as $N\rightarrow \infty$ in the AF phase.
It also follows immediately from the 
the law of large numbers that, in the paramagnetic phase, $\Delta_m^2\rightarrow \pi /2 -1$ as $N\rightarrow \infty$.  These two statements imply that curves of $\Delta_m^2$ vs $T$ for various values of $N$ cross at the phase boundary between the paramagnetic and AF phases. We make use of this fact to quantitatively determine the AF-paramagnet phase boundary.
The same criterion can be applied to the AF-SG phase boundary. To see why this is so, note that, the plots shown in Fig. \ref{3magne}b for $x=0.5$ suggest 
$m_2\to N^{-1}$ as $N\to \infty$, even at low temperatures, that is, well within the SG phase.
Plots of $\Delta_m^2$ vs $T$ are shown
in Figs. \ref{4d2xalto}a and \ref{4d2xalto}b for $x=0.7$ and $0.6$,  respectively.
The signature of an AF phase below $T \simeq 1.2$ clearly shows up in Fig. \ref{4d2xalto}a. We have thus established all points of the AF phase boundary
shown in Fig. \ref{1fases} for $x\geq 0.7$. For the low temperature portion of the phase boundary (near $x=0.65$) this procedure is not very effective.
From Fig. \ref{4d2xalto}b, we infer that the AF boundary line must drop to a $T=0$ value at some $x>0.60$. The three data points shown for $x\simeq 0.65$ and $T<1$
are obtained from plots such as the one shown in the inset of Fig. \ref{1fases} for $T=0.4$.

\section{The SG phase}
\label{resultsB}

In this section, we report numerical results we draw from tempered MC calculations
for $q_2$, for distributions of $q$, and for $\xi_L$.
Because we expect, from the argument given in Sec. I, lattice independent behavior for $x\ll 1$, we emphasize
the results we have obtained for the two smallest values of $x$ we have dealt with, $x=0.2$ and $x=0.35$ (that is,
$x\simeq 0.3x_c$ and $x\simeq 0.54x_c$).

\begin{figure}[!t]
\includegraphics*[width=75mm]{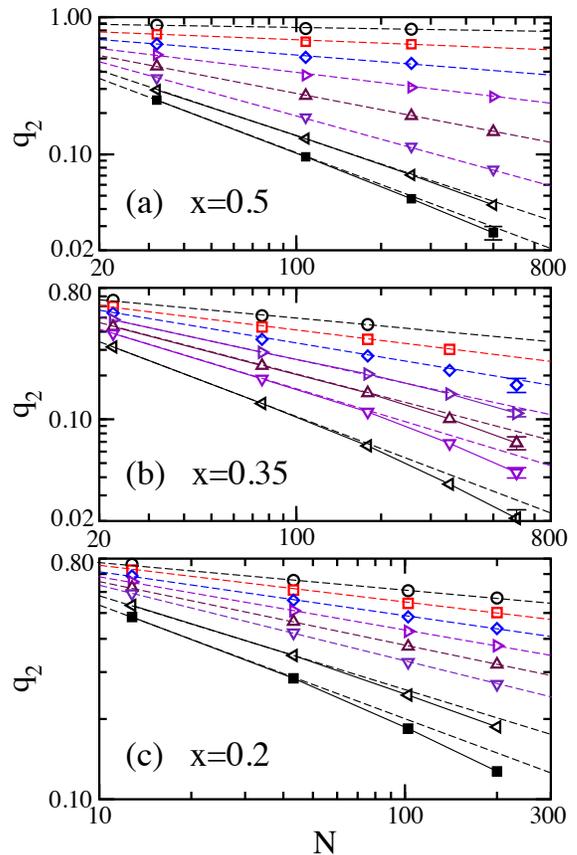}
\caption{ (Color online)
(a) Plots of $q_2$ versus the number of dipoles $N$ for $x=0.5$. 
$\circ$, $\square$, $\diamond$, $\triangleright$, $\triangle$, $\triangledown$, $\triangleleft$, and
$\blacksquare$ stand for $T=0.1, 0.2, 0.3, 0.4, 0.5, 0.6, 0.7$, and $0.8$ respectively.
Lines are guides to the eyes.
(b) Same as in (a) but for $x=0.35$.
$\circ$, $\square$, $\diamond$, $\triangleright$, $\triangle$, $\triangledown$, and $\triangleleft$, 
stand for $T=0.25, 0.3, 0.35, 0.4, 0.45, 0.5$, and $0.6$ respectively.
(c) Same as in (a) but for $x=0.2$.
$\circ$, $\square$, $\diamond$, $\triangleright$, $\triangle$, $\triangledown$, $\triangleleft$, and
$\blacksquare$,  stand for $T=0.12, 0.14, 0.16,..., 0.22, 0.26$, and $0.30$.
For all data, we have checked that, within errors,  $\tilde q_2=q_2$.
 Clearly, data point sets for larger temperatures deviate from
the straight dashed-lines shown (implying faster than a power of $1/L$ decay) while sets for lower temperatures do not. 
Error bars are shown only where they are larger than the icon sizes.  For each set of points with given $x$ and $T$ values, 
$\chi^2$ values for straight line fits, as well as
the largest error, are given in Table III. } 
\label{6q2vsN}
\end{figure}

A plot of  $q_2$ versus $T$ is shown in Fig.~\ref{5q2vsT}.
Note that  $q_2$ decreases as $N$ increases, even at low temperatures. We have found similar behavior
 for other values of $x$ satisfying  $x\lesssim x_c$.  Inspection of this figure raises the question
of whether $q_2$ vanishes as $L\rightarrow \infty$. In order to advance in this direction, we do
log-log plots of $q_2$ vs $N$, which we show in  Figs. \ref{6q2vsN}a,  \ref{6q2vsN}b, and \ref{6q2vsN}c,
for the values of $x$ shown therein. The data points in these three figures seem consistent with,
$q_2\sim N^{-p}$ 
for $T/x\lesssim 1$, where $3p=1+\eta$, as follows from the definition of $\eta$ in Sec. \ref{wlro} (see also Ref. \onlinecite{mefisher}). 
 $\chi^2$ values for $q_2\sim N^{-p}$ fits to sets of data points, for $T/x\lesssim 1$ (for which they are appropriate) as well as for $T/x\gtrsim 1$ (for which they are not appropriate), are given in Table III. 
Plots of  $q_1$  vs $N$ show the same qualitative behavior.
All of this is in accordance with quasi-long-range order. We return to this point below and in Sec. \ref{wlro}.

\begin{table}\footnotesize
\caption{   $\chi^2_r$ values for two-parameter $q_2=c/N^p$ fits to sets of data points for $q_2$ vs  $T$ displayed in Figs. \ref{6q2vsN}a-c.
As usual, we define $\chi_r^2=\chi^2/df$, where $df$ is the number of data points in each set minus the number of fitting parameters (2, here). 
The largest errors $\Delta q_2$ of $q_2$ from all data points for each $x$ and $T$ are also given.} 
\begin{ruledtabular}
\begin{tabular}{ p{2.0cm} p{2.0cm} p{2.0cm}}
  $x=0.50$   & $x=0.35$  &   $x=0.20$   \\
\end{tabular}
\begin{tabular}{ {c}{c}{c}{c}{c}{c}{c}{c}{c}}
  $T$& $\chi^2_r$&  $\Delta q_2$&   $T$& $\chi^2_r$&  $\Delta q_2$&   $T$& $\chi^2_r$&  $\Delta q_2$  \\  
\end{tabular}
\begin{tabular}{ |c|c|c|c|c|c|c|c|c|}
0.10&1.29&0.01&   0.20&0.21&0.008&   0.12&0.28&0.01\\
 0.20&0.84&0.01&   0.30&0.70&0.01&    0.14&0.22&0.01\\
 0.30&0.91&0.01&   0.35&0.38&0.02&    0.16&0.15&0.01\\
 0.40&0.96&0.008&  0.40&0.52&0.012&   0.18&0.08&0.01 \\
 0.50&0.12&0.006&  0.45&1.70&0.008&   0.20&0.03&0.01\\
 0.60&0.46&0.004&  0.50&3.50&0.004&   0.22&0.12&0.01\\
 0.70&1.96&0.004&  0.60&15.09&0.003&  0.26&1.24&0.008\\
 0.80&2.20&0.003&      &     &     &  0.30&3.38&0.006\\
\end{tabular}
\label{table}
\end{ruledtabular}
\end{table}

Reading off values of $p$ from plots shown in Figs. \ref{6q2vsN}a,  \ref{6q2vsN}b, and \ref{6q2vsN}c,
we obtain $\eta$  for $x \le 0.5$ and various values of $T$. The relation 
$\eta =-1+a_x(T/x)^2$ fits the data  rather well for all $T/x\lesssim 1$, if we let $a_x=0.76, \;0.98,\; 1.18$ for $x=0.2, \;0.35, \; 0.5$, respectively. In order to be able to conclude that $\eta (T_{sg})$ varies with $x$, we would need to know $T_{sg}$ within an error of $10\%$. 
Unfortunately, we find below (in Sec. \ref{valueTsg}) an error in $T_{sg}$ which is not much smaller than $10\%$.

For higher values of $T/x$, $q_2$ vs $N$  curves downwards, as expected for
the paramagnetic phase. Approximate values of $T_{sg}$ can thus be obtained from such plots, but
more accurate methods are given below. 
It is reassuring to see in Figs. \ref{6q2vsN}a,  \ref{6q2vsN}b and \ref{6q2vsN}c,
the values of  $\tilde q_2$ we have obtained agree, within errors,
with the values for $q_2$. 

\begin{figure}[!t]
\includegraphics*[width=73mm]{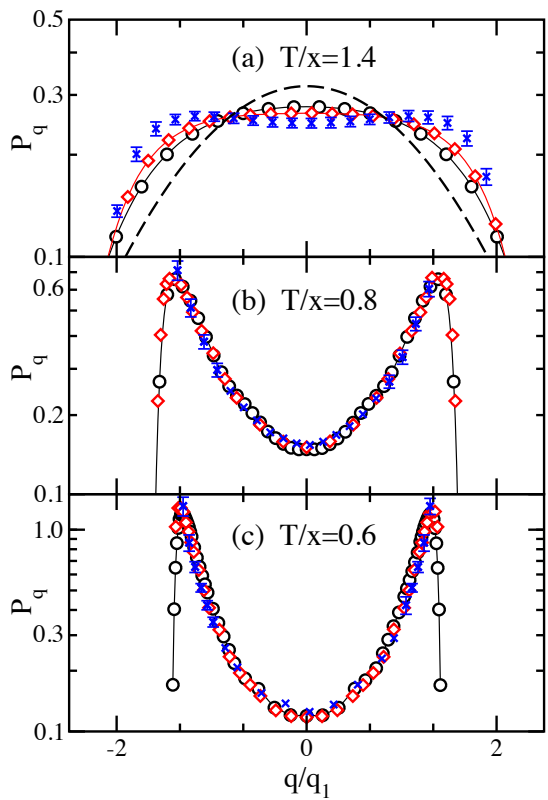}
\caption{ (Color online)
(a) Plots of the probability distribution $P_q$ versus $q/q_1$ for $x=0.2$ and $T/x=0.4$. 
$\circ$, $\square$, $\times$  are for $L=10, 8$ and $6$, respectively.  
The thick dashed line is for the Gaussian distribution
that ensues for a paramagnet in the
macroscopic limit.
(b) Same as in (a) but for $T=0.16$. 
(c) Same as in (a) but for $T=0.12$.
Error bars are shown wherever they are larger than symbol sizes.}

\label{7pdqus}
\end{figure}

We next give distributions of $q$ we have found. We make use of a normalized distribution 
$P_q(q_r)$, where $q_r=q/q_1$. In macroscopic paramagnets, $q_r$ is expected to be normally distributed, 
as follows from the law of large numbers and the fact that spin-spin correlation lengths are then finite.
On the other hand, $P_q=[\delta(q_r-1)+\delta(q_r-1)]/2$, where $\delta$ is the Dirac delta function, in a SG phase,
according to the droplet picture of SGs.\cite{droplet}
Plots of $P_q$ vs $q_r$ are shown  for $x=0.2$ in 
Figs. \ref{7pdqus}a,  \ref{7pdqus}b, and \ref{7pdqus}c.
Clearly, $P_q(q_r)$ drifts with system size in Fig. \ref{7pdqus}a, for $T=0.28$.
Our results are consistent 
with  $P_q(q_r)\rightarrow (1/\pi )\exp (-q_r^2/\pi )$ as 
$N\rightarrow \infty$, which is in accordance with a paramagnetic phase. 
On the other hand, we find for lower temperatures double peaked distributions in Figs. \ref{7pdqus}b and Fig. \ref{7pdqus}c that are fairly broad  
and, within errors, do not change with $N$. 
This is contrary to the prediction of the droplet-model theory of SGs. 
From these graphs we conclude that $0.16<T_{sg}<0.26$ for
$x=0.2$. Analogous plots for $x=0.35$ (not shown) give $0.30<T_{sg}<0.45$.

\begin{figure}[!t]
\includegraphics*[width=87mm]{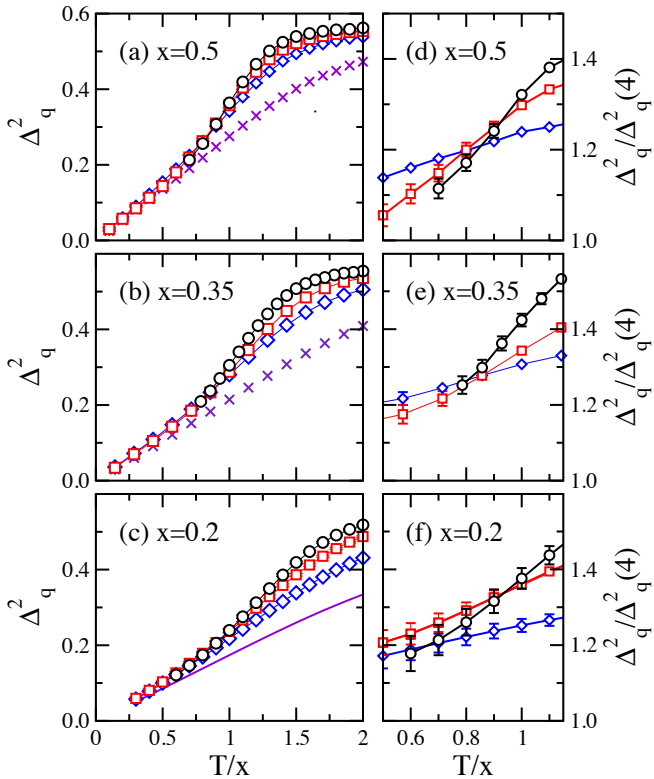}
\caption{(Color online) (a) Plots of  $\Delta ^2_q$ vs $T/x$, for $x=0.5$.
$\circ$, $\square$, $\diamond$,  and $\times$ are for $L=10, 8, 6$ and $4$, respectively.
(b) Same as in (a) but for $x=0.35$.  (c)  Same as in (a) but for $x=0.2$.
(d) Plots of  $\Delta ^2_q/\Delta ^2_q(4) $ vs $T$ for $x=0.5$. Symbols are as in (a).
(e) Same as in (d) but for $x=0.35$.
(f) Same as in (d) but for $x=0.2$.  In panels (a), (b), and (c), all error bars are smaller than symbol sizes.}
\label{8d2qvsTx}
\end{figure}

Results for  the scale free quantity $\Delta_q^2$ follow. Recall that, 
as explained for  $\Delta_m^2$, $\Delta_q^2 \to \pi/2-1$
as $N\to \infty$ in the paramagnetic phase, vanishes when there is strong long-range 
order, and goes,  at
the critical temperature, to some intermediate value that is size independent.
 This is as shown in Fig. \ref{4d2xalto}c for $x=0.7$ where 
curves for various values of $N$ 
cross at $T_{AF}$.    
Figures \ref{4d2xalto}a and  \ref{4d2xalto}c look rather similar, because $q$ and $m$ are not qualitatively different in the
AF phase. This is not so for $x<x_{c}$, where there is no AF order. 
Figures \ref{4d2xalto}b and \ref{4d2xalto}c for $x=0.6$ show that, within errors, curves 
of $\Delta_q^2$ vs $T$ for different system sizes merge (not 
cross) near $T=0.65$, while $\Delta_m^2$
increases with $N$ for all temperatures. Similarly,  
$\Delta_q^2$ vs $T$ curves merge, for $x=0.65$, near $T=0.75$ (not shown). 
Plots of $\Delta_q^2$ vs $T/x$ are shown in 
Figs. \ref{8d2qvsTx}a, \ref{8d2qvsTx}b, and \ref{8d2qvsTx}c  for lower concentrations.

We notice that curves in Figs. \ref{8d2qvsTx}a, \ref{8d2qvsTx}b, and \ref{8d2qvsTx}c  differ only slightly. 
This follows from the argument given in Sec. I, which shows that 
all physical quantities for three dimensional dipolar systems can only be functions of $T/x$ for $x\ll 1$. 
The data points in Fig. \ref{8d2qvsTx} show that $\Delta_q^2\rightarrow \pi /2-1$ 
as $N\rightarrow\infty$, for $T/x\gtrsim 1$, as expected for the paramagnetic phase.

Curves for $\Delta_q^2$ vs $T$ seem to merge at
a lower temperature, near $T/x=0.9$. However, closer scrutiny shows that these curves actually cross, 
albeit at very small glancing angles. This can be appreciated in Figs.  \ref{8d2qvsTx}d, \ref{8d2qvsTx}e, and \ref{8d2qvsTx}f,
where plots of the ratios $\Delta_q^2(L)/\Delta_q^2(4)$ vs. $T$ are given for various values of $L$, for $x=0.5$, $x=0.35$, and $x=0.2$, respectively.
Note that the weak dependence of $\Delta_q^2$ with system size at low temperatures
is in accordance with our result that $P_q(q_r)$  does not change appreciably with system size below $T_{sg}$. This point is further elaborated
in Sec. \ref{wlro}

\begin{figure}[!h]
\includegraphics*[width=70mm]{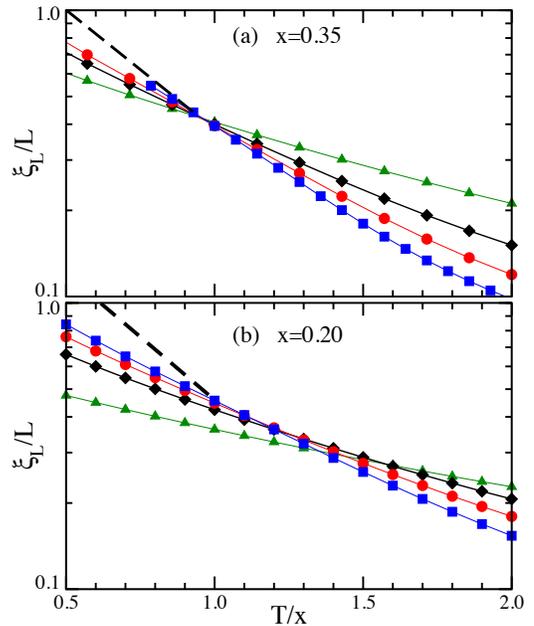}
\caption{(Color online)
(a) Semilog plots of (a) $\xi_L /L$
versus $T/x$ for $x=0.35$, and 
$L=10$ ($\blacksquare$), $L=8$ ($\bullet$), $L=6$ ($\blacklozenge$),  and $L=4$ ($\blacktriangle$). 
Dashed line follows from $1/L\to 0$ straight line extrapolations in the plots shown in Fig. \ref{10xivsN}a for $T<T_{sg}$. Continuous lines are guides to the eye.
(b) Same as in (a) but for $x=0.2$. All error bars are smaller than symbol sizes.}
\label{9longitudes}
\end{figure}

Following the lead of Refs. \onlinecite{longi} and \onlinecite{balle}, who have found that $\xi_L/L$ (defined in
Sec. \ref{meas}) crosses at $T_{sg}$ and spreads out as $T$ decreases below $T_{sg}$ for the EA model 
in 3D, we next examine how  $\xi_L/L$ behaves for the PAD model. As pointed out in Sec. I and  Table I, this
has already been done for the PAD model on a LiHo$_x$Y$_{1-x}$Y$_4$ lattice by Kam and Gingras.\cite{gin} As we also
point out in Sec. I, we aim to explore the behavior of the PAD model, not only near $T_{sg}$, but also deep into the SG phase.
Recall that $\xi_L$ becomes a true correlation 
length when $\xi _L/L\ll 1$. Then, in the paramagnetic phase, $\xi_L/L\sim O(1/L)$, therefore decreasing as $L$ increases.
At $T=T_{sg}$, $\xi_L/L$ must become size independent, as expected 
for a scale free quantity.
The inferences one can make about the nature of the condensed phase from the behavior of $\xi_L$ where $T<T_{sg}$ is the
subject of Sec. \ref{wlro}. Without further comment, we next report our results.
Plots of $\xi_L/L$ versus  $T/x$ are shown in Figs. \ref{9longitudes}a and \ref{9longitudes}b for 
$x=0.35$ and $0.2$, respectively.  
Note that curves spread out above and below $T/x\sim 1$. 
For $x=0.35$, curves for all $L$ cross at $T_{sg}/x=0.95(5)$. 
On the other hand, the temperatures where pairs of curves 
for lengths $L_2$ and $L_1$ cross for $x=0.2$ 
decrease as lengths $L_2$ and $L_1$ increase (see Fig. \ref{9longitudes}b), pointing to a $T_{sg}/x\lesssim 1.1$.

\begin{figure}[!t]
\includegraphics*[width=83mm]{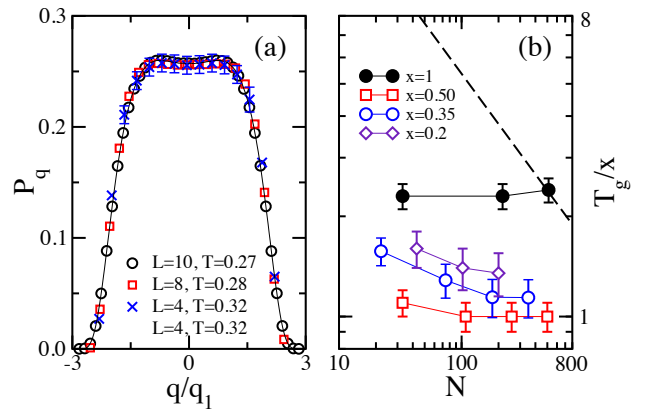}
\caption{(Color online)
(a) Plots of distributions $P_q$ versus $q/q_1$ for $x=0.2$ and the shown values 
of $L$ and $T$.  Error bars are shown only where they are larger than symbol sizes.
(b) Plots of $T_g/x$ versus $N$ for the shown values of $x$.
The thick dashed line stands for the $N^{-1/2}$ behavior obtained in Ref. \onlinecite{yu}.
}
\label{yua}
\end{figure}

\section{Existence and nature of the SG phase}
\label{character}

In this section we examine the numerical results given in the previous section.
We (i) arrive at values for $T_{sg}$ as a function of $x$, (ii) show that weak long-range order is consistent with our results for the SG phase, and (iii)
draw values for the critical exponent $\nu$ for various values of $x$.

\subsection{The value of $T_{sg}$}
\label{valueTsg}

Recall first that $\Delta_q^2$ vs $T$ curves for different values of $L$ are supposed to come together as
$T$ approaches $T_{sg}$ from above. This behavior is exhibited in Figs. \ref{8d2qvsTx}a-c. A closer view
of how such curves actually meet at $T=T_{sg}$ is offered in Figs. \ref{8d2qvsTx}d-f, where plots of $y(L,4)$ versus $T/x$,
where $y(L,L^\prime)=\Delta^2_q(L)/\Delta^2_q(L')$, are shown. One aims to find the $L\to\infty$ and $L'\to\infty$ limit
of $y(L,L')=1$, which gives the value of $T_{sg}$.  We find that
$y(L,L')=1$ at values of $T/x$ that increase  with $L$ and $L'$, which is reassuring, because it shows that
$T_{sg}$ does not vanish. Furthermore, we draw the following lower bounds from the plots in Figs. \ref{8d2qvsTx}d-f,
$T_{sg}/x\gtrsim 0.95,\; 0.8,\;0.95$, for $x=0.5,\; 0.35,\;0.20$, respectively.

We obtain a complementary determination of $T_{sg}$ from the intersection of $\xi_L/L$ vs $T$ curves.
This is as is sometimes done for the EA\cite{longi,balle,katz0} and PAD\cite{gin} models.
We obtain, from Fig. \ref{9longitudes}a, $T_{sg}/x\simeq 0.95$ for $x=0.35$. In Fig. \ref{9longitudes}b, we see that
$\xi_L/L$ vs $T$ curves meet at decreasingly smaller values of $T$ as $L$ increases. 
We thus obtain $T_{sg}/x\lesssim 1.1$ for $x=0.2$. 

From these two complementary determinations, we arrive at: $T_{sg}/x=1.0(1)$ for $x\lesssim 0.5$.

An aside follows about the result  by Snider and Yu,\cite{yu} that $T_{sg}=0$ for $x=0.045, 0.12$ or $0.2$.
This is, of course, in clear contradiction with our results. Their conclusions come from their work with the Wang-Landau\cite{wl} variation of the MC algorithm. 
Their evidence is from plots of $T_g$ versus $N$, where $T_g$ is the temperature 
at which $P_q$ becomes flattest. This procedure makes sense because  $T_g \to T_{sg}$
as $N\to\infty$. They found $T_g$ to vanish as $N^{-1/2}$ for several $x$ values, including $x=0.2$. 
We now repeat this procedure using our own data, including the ones for $x=0.2$.
In Fig. \ref{yua}a we plot the flattest distributions we found for $x=0.2$ and $L=4,8$, and $10$.  Note in passing that all 
scaled distributions coincide and have therefore the same value  of $\Delta^2_q$. 
Plots of the values of $T_g/x$ we have obtained for $x=0.5,\;0.35$, and $0.2$ are shown in Fig. \ref{yua}b. 
Our data points are in clear contrast to the $T_g\sim N^{-1/2}$ trend of Ref. \onlinecite{yu}, and
point to $T_{sg}/x\simeq 1$.  Whether this disagreement comes from using a different
Monte Carlo method, or from the unusual definition of $q$ in Ref.  \onlinecite{yu}, we do not know. 

\subsection{Marginal behavior}
\label{wlro}

Here we discuss how various pieces of evidence (including crossings of $\xi_L/L$ vs $T$ curves) lead us to the conclusion that the SG phase of the PAD model behaves marginally.
That is to say, that $\langle q^2\rangle\to 0$ and $\chi_{sg}\to \infty$ in the macroscopic limit.

The variation of $\langle q^2\rangle$ with $L$ for various temperatures, exhibited in Figs. \ref{6q2vsN}a-c,
has already been considered in Sec. \ref{resultsB}. For all $x<x_c$, $T<T_{sg}$, and all system sizes we have studied, 
we find no deviation from $\langle q^2\rangle \sim L^{-(1+\eta)}$. Nor do we find any size dependence
in $P_q(q_r)$.  This is illustrated in Figs. \ref{7pdqus}b and c,
and is in accordance with the behavior of the distribution of the magnetization that is observed\cite{rad2}  in the condensed phase of the 2D $XY$ model.
Note that the variation of $\Delta_q^2$ with system size is a measure of the variation of $P_q(q_r)$. 
The very small changes we have observed in $\Delta_q^2$ as $L$ varies in the PAD model for all $T\lesssim T_{sg}$ turn out to be smaller than 
the corresponding changes in the $XY$ model.\cite{rad2} This is, of course, in marked contrast with the behavior one expects of the corresponding quantity for a strongly ordered system, such as the 
droplet model of SGs or an ordinary ferromagnet, in which $\Delta_q^2\to 0$ in the 
macroscopic limit of the ordered phase.  Neither do our results fit into a RSB 
scenario,\cite{RSB} in which 
$q_2$ does not vanish as $L \to \infty$ and would have $P_q(q_r)$ changing with system size,
since $P_q(q)$ is wide and does not change with system size in the SG phase.

\begin{figure}[!t]
\includegraphics*[width=75mm]{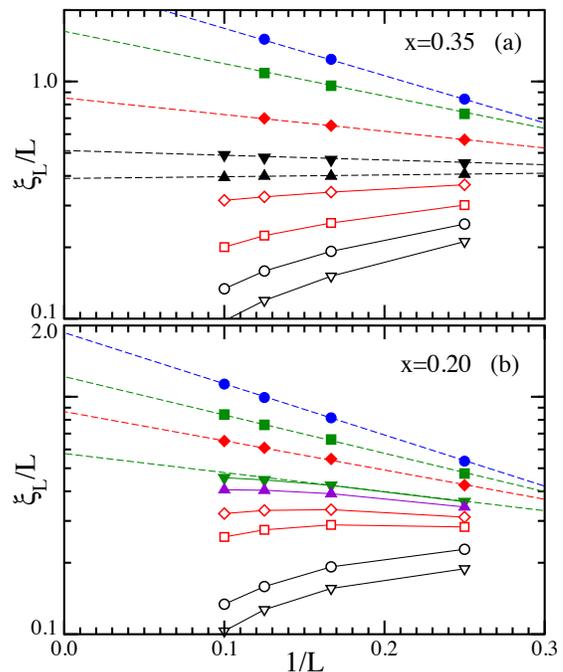}
\caption{(Color online)
(a) Semilog plots of $\xi_L/L$ versus $1/L$ for $x=0.35$, and  
$T/x=0.143$ ($\bullet$), $T/x=0.286$ ($\blacksquare$), $T/x=0.571$ ($\blacklozenge$), $T/x=0.857$ ($\blacktriangledown$), $T/x=1.00$($\blacktriangle$),
 $T/x=1.14$ ($\lozenge$), $T/x=1.43$ ($\square$), $T/x=1.71$ ($\circ$), and $T/x=2.00$ ($\triangledown$). 
 (b) Same as in (a), but for  for $x=0.20$, and  
$T/x=0.300$ ($\bullet$), $T/x=0.500$ ($\blacksquare$), $T/x=0.700$ ($\blacklozenge$), $T/x=1.00$ ($\blacktriangledown$), $T/x=1.10$($\blacktriangle$),
 $T/x=1.30$ ($\lozenge$), $T/x=1.50$ ($\square$), $T/x=2.00$ ($\circ$), and $T/x=2.50$ ($\triangledown$).  
 All errors are: between $2\%$ and $3\%$  in (a), and between $2\%$ and $4\%$ in (b), and are thus hidden behind the icons. In both (a) and (b), the straight-dashed lines give $\chi_r^2 <1$ fitting values, except for $T/x=1.0$ in (b), for which $\chi_r^2 =3.3$ } 
\label{10xivsN}
\end{figure}

We now analyze the data we have for $\xi_L$. First, we outline how we expect $\xi_L/L$ to spread out
as $T$ decreases below $T_{sg}$ in various SG scenarios.

(i) \emph{Condensate with short range order fluctuations}.
In such a SG phase, $q_2\neq 0$ and $\langle \phi_0\phi_r \rangle -\langle \phi_0\rangle  \langle\phi_r \rangle$ would be short ranged. 
This would fit into the droplet model of spin glasses.\cite{droplet}
It then follows straightforwardly from its definition [Eq. (\ref {phi1})] { that $\xi_L^2/L^2\sim L^d$. }Here, $d=3$, and there is nothing in the
plots of $\xi_L/L$ vs $1/L$, which are shown in Figs. \ref{10xivsN}a and  \ref{10xivsN}b, to suggest that { $\xi_L^2/L^2\sim L^3$ } at any nonzero temperature. 

(ii) \emph{Condensate with long range order fluctuations}.

Let $\langle A \rangle_q$ be the thermal average of $A$ over all states with a given $q$ value. Clearly, $\langle A \rangle =\int \langle A \rangle_q P_qdq $.
{ Assume $q_2\neq 0$, and $\int [ \langle \phi_0\phi_r \rangle_q -q^2 ] P_q dq=G(r)$, where,}
\begin{equation}
G(r)\equiv \frac {A}{r^{d-2+\eta}},
\label{G}
\end{equation}
for $r\gg a$, where $A$ is a constant. This behavior fits in with the RSB picture.\cite{RSB}
Then, it follows from its definition [Eq. (\ref {phi1})] that { $\xi_L^2/L^2\sim L^{1+\eta}$. } Recall, from Sec. \ref{resultsB},
that $\eta\simeq -1+(T/T_{sg})^2$ in the SG phase. 
Evidence for { $\xi_L^2/L^2\sim L^{1+\eta}$ } appears neither in Fig. \ref{10xivsN}a nor in Fig. \ref{10xivsN}b.

(iii) \emph{Marginal behavior}.
Then, $q_2 = 0$ and $\langle \phi_0\phi_r \rangle =G(r)$. This is as in the KT theory\cite{xy}
of the 2D $XY$ model. It then follows straightforwardly from the definition of $\xi_L/L$ that $\xi_L/L$ becomes
independent of $L$ for very large $L$. This is precisely the outcome from $1/L\to 0$ extrapolations of the straight lines shown in Fig. \ref{10xivsN}a and \ref{10xivsN}b
for all $T/x\lesssim 1$. 

\begin{figure}[!t]
\includegraphics*[width=75mm]{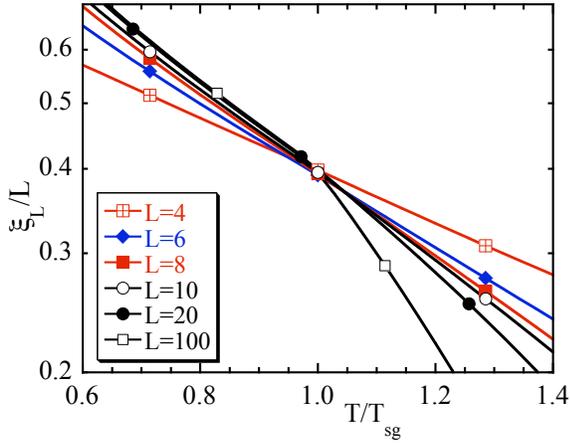}
\caption{(Color online) Semilog plots of $\xi_L/L$ vs $T/x$ from Eq. (\ref{G}) for the shown values of $L$.
In Eq. (\ref{G}), we let $A=0.67$, and $\eta =-1+(T/T_{sg})^2$. }
\label{nueva}
\end{figure}

Note also in Figs. 12a and 12b that curves for $\xi_L/L$ vs $1/L$ become steeper as $T$ decreases below $T/x\simeq 1$. Now, recall from above that $q_2\neq 0$ implies $\xi_L^2/L^2\sim L^d$ and $\xi_L^2/L^2\sim L^{1+\eta}$, for short- and long-range fluctuations from the condensate. Note further that $\mid 1+\eta \mid$ decreases as $T$ decreases. This would lead to $\xi_L/L$ vs $1/L$ curves which do not become steeper as $T$ decreases below $T/x\simeq 1$, which is in clear contradiction with the observed behavior. This is an additional piece of evidence for quasi long-range order.

Thus, the most straightforward interpretation of the data shown in Figs.  \ref{10xivsN}a and \ref{10xivsN}b leads us to suspect that the SG phase in the PAD model behaves marginally. 
This might seem to be in contradiction to the fact that $\xi_L/L$ curves do cross, as shown in Fig. \ref{9longitudes}, and that, as pointed out in Ref. \onlinecite{balle},
$\xi_L/L$ vs $T$ curves merge, not cross, for the 2D XY model, as $T\to T_{sg}$ from above. (Indeed, no crossings occur for even much smaller 2D XY systems than the ones for which data points are shown in Ref. \onlinecite{balle}). We next give a specific example in order to illustrate how both merging and spreading out as $T$ decreases below $T_{sg}$ can take place, depending on the some details in $G(r)$.

\begin{figure}[!t]
\includegraphics*[width=75mm]{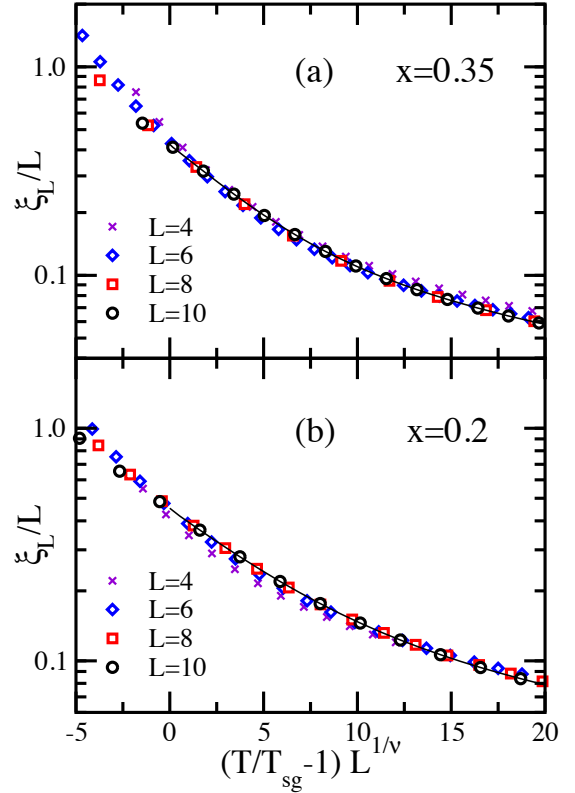}
\caption{(Color online)
(a) Semilog plots of $\xi_L/L$ versus $(T/T_{sg}-1)L^{1/\nu}$ for $x=0.35$, $T_{sg}=0.345$, $\nu =0.95$, and the shown values of $L$.
(b) Same as in (a) but for $x=0.20$,  $T_{sg}=0.21$, $\nu =0.95$, and the shown values of $L$. 
 Recall that scaling is expected only for $T/T_{sg}-1>0$. In both panels, all error bars are somewhat smaller than the icon sizes. }
\label{scaleXi}
\end{figure}

We first calculate $\xi_L/L$ from  $\langle \phi_0\phi_r \rangle =G(r)$ and Eq. (\ref{G}) for all $r$ except
that $G(r)=1$ for all $r\leq 1$. 
To proceed, we let $A=0.67$ for $T\leq T_{sg}$ but  not too close to  $T=0$, where one expects $A=1$. 
We are not interested here in the $T>T_{sg}$ range, but we nevertheless then let $A\to Ae^{-r/\xi_\infty}$, $\xi_\infty=7(T/T_{sg}-1)^{-\nu} $, 
and $\nu=1$, which is roughly the value we obtain below  (see Sect. \ref{nuexp}). We 
make use of $\eta =-1+(T/T_{sg})^2$, 
which we have found in Sec. \ref{resultsB}. Finally, in order to be able to
make comparisons with our MC results, which we have obtained for periodic boundary conditions,  
we let in Eq. (\ref{G}),
\begin{equation}
r\to Q^{-1}[ \sum_{\alpha=1}^{3} \sin^2(Q r_{\alpha}  ) ]^{1/2},
\end{equation}
where $Q=\pi /L$ and $\textbf {r}=(r_1,r_2,r_3)$.
Straightforward numerical implementation of Eq. (\ref{phi1}) yields the data points that 
are plotted in Fig. \ref{nueva}.  Note the resemblance between
Fig. \ref{nueva} and Figs. \ref{9longitudes}a and \ref{9longitudes}b 
which follow from our MC calculations.

Merging of $\xi_{L}/L$ curves at $T=T_{sg}$ as $T$ decreases is obtained for all $L\geq 4$ if, instead of $A=0.667$, we let $3A=3-(T/T_{sg})^2$.
Note that $A(T_{sg})=0.667$ and $A(0)=1$. If, on the other hand, one lets $3A=3-(T/T_{sg})^s$ and  $0<s\lesssim 0.2$, which satisfies the same end point conditions, 
one obtains plots for $\xi_L/L$ vs $T$ which look much like the ones shown in Figs. \ref{9longitudes}. 

To summarize, all our data (including spreading out of $\xi_L/L$ curves as $T$ 
decreases below $T_{sg}$) are consistent with marginal behavior in which 
the correlation length diverges at $T_{sg}$ as in a conventional phase transition,
but weak-long-range order occurs  below $T_{sg}$, as in the 2D XY model.

\subsection{The $\nu$ exponent}
\label{nuexp}

In accordance with the above, we look for the values of $\nu$ and $T_{sg}$ which best collapse $\xi_L/L$ vs $(T/T_{sg}-1)L^{1/\nu}$ plots for various values of $L$ into a single curve for temperatures above $T_{sg}$. 
The best results, exhibited in Figs.  \ref{scaleXi}a and  \ref{scaleXi}b, for  $x=0.35$ and $x=0.20$, 
are obtained with $T_{sg}/x=1.0(1)$ and $\nu=0.95$.
 Note the data points scatter below $T_{sg}$. This is as expected, and is consistent with quasi-long range  order in the SG phase, since $\xi_L/L$ becomes independent of $L$ then for sufficiently large $L$. 
Note that, as in the  EA model,\cite{katz0} $L=4$  seems to be too small to scale properly.

\section{DISCUSSION}
\label{discussion}

By tempered Monte Carlo calculations, we have studied an Ising model on a simple cubic lattice. There are only dipole-dipole interactions.
Spins (randomly) occupy only a fraction $x$ of all lattice sites.
We have calculated the entire phase diagram of the system. It is shown in Fig. \ref{1fases}.
We have also provided strong evidence for the existence a SG phase for $0<x<x_c$, where $x_c= 0.65(5)$.
The SG transition temperature is given by $T_{sg}(x) \simeq x$.  We have argued in Sec. I that 
this result carries over into other lattices if (i) $x\ll 1$, and (ii) we replace the latter expression for $T_{sg}$ by 
$k_BT_{sg}=n_d\varepsilon_d$ ( see Table I). How we have arrived a this conclusion is described in Sec. \ref {valueTsg}.

We have not dwelt on the applicability of our MC results to experiments.
That is beyond the scope of this paper. We nevertheless make a few comments.
Recall first that, as we argue in Sec. I, lattice structure is of no consequence
for very dilute PAD models. Then, $T_{sg}$ as well as the temperature $T_m$ where the
specific heat takes its maximum value can only depend (as in the MC simulations of Ref. \onlinecite{bh}) on $n_d\varepsilon_d$ (see Table I).
We notice in { Table I }  values for $T_{sg}$ do not fully comply with this rule.  
In addition, in very dilute LiHo$_x$Y$_{1-x}$F$_4$ systems, $T_m$ hardly changes with $x$. \cite{quilliam}  
There are several sources for the discrepancies between experiments on very dilute LiHo$_x$Y$_{1-x}$F$_4$ and
the PAD model. 
Quantum effects seem to play a role in experiments on very dilute  LiHo$_x$Y$_{1-x}$F$_4$ systems.\cite{gosh}
This is not too surprising, since tunneling can become relevant when barrier energies become overwhelmingly large.
However, we do not expect small perturbations that bring about tunneling and concomitant time dependent effects to have a significant effect on \emph{equilibrium} properties, which is the subject of this paper. In addition, exchange couplings among nearest neighbor spins\cite{bh,nj} are disregarded in the PAD model we study. Note, however that the effect of nearest neighbor interactions must vanish as  $x\to 0$. {\it Clustering} of the spatial distribution of dipoles can also lead to discrepancies.\cite{gosh}
None of the above can however account for (i) the numerical differences between the MC results (see { Table I }) of Tam and Gingras,\cite{gin} and ours, nor can they account for the more serious discrepancy with (ii) Ref. \onlinecite{yu}, which we discuss in some detail in Sec. \ref{valueTsg}. Numerical (not too large) discrepancies notwithstanding, our results support the ones from 
Tam and Gingras,\cite{gin} that the dilute PAD model does have a SG phase. On the other hand, for the roots of the discrepancies with experimental results (see { Table I }) on dilute LiHo$_x$Y$_{1-x}$F$_4$ systems, we have no clear picture.

As for the nature of the SG phase, all of our results are consistent with quasi-long-range order. Full details are given in Sec. \ref{wlro}. We know of no previous study of the nature of the SG phase of the PAD model with which to compare our results. (Only the critical behavior of a PAD model is examined in Ref. \onlinecite{gin}.) On the other hand, our conclusion for the PAD model can be compared with and one drawn  for the EA model in Refs. \onlinecite{longi,balle,katz0}. They are both based on the behavior of $\xi_L/L$ vs $T$ curves for various values of $L$. The conclusions differ, not so much because of the data, but because we have looked at the data differently (see Sec. \ref{wlro} and Refs. \onlinecite{longi,balle,katz0}).

\acknowledgments
For different helpful comments, we are grateful to Prof. Amnon Aharony, Prof. Michael E. Fisher, and Prof. Jacques Villain. We are specially indebted to Prof. JV for kindly reading the manuscript.
We are indebted to the Centro de Supercomputaci\'on y Bioinform\'atica and to the 
Applied Mathematics Department both at University of M\'alaga, and to 
Institute Carlos I at University of Granada  for much computer time. Finally, we thank 
financial support from Grant FIS2006-00708
from the Ministerio de Ciencia e Innovaci\'on of Spain.

{
\appendix
\section{WHY WE DO NOT DO EWALD SUMS}

We consider site-diluted systems of Ising magnetic dipoles in a cubic box of $L^3$ sites on a SC lattice. 
All dipoles point along the $z$ axis of the lattice. Each site is occupied with probability $x$.  
We assume thermal equilibrium. We show two things in this appendix. We first show that the contribution $\Delta h$ to the magnetic field $h$ at the center of such box, coming from a periodic arrangement of replicas that span all space beyond the system of interest (the ``outer space'') within an arbitrarily large cube which is centered on the system of interest, vanishes as $L\rightarrow \infty$ if the system is not in a ferromagnetic phase or close to its Curie temperature. More specifically, we show that if $\langle s_is_j \rangle -\langle s_i \rangle \langle s_j \rangle$ is short ranged, and the system is homogeneous (including antiferromagnetically ordered states), then
\begin{equation}
\langle {\Delta h^2} \rangle \rightarrow 0
\label{aa1}
\end{equation}
as $L\rightarrow\infty$, where $\langle \ldots \rangle$ stands for an average over both a canonical ensemble and (site occupation) disorder. Note that we are not imposing the condition that $\langle s_is_j \rangle ^2- \langle s_i \rangle ^2 \langle s_j \rangle ^2 $ be short ranged, and recall (1) that in general $\sum_j\langle s_is_j \rangle - \langle s_i \rangle \langle s_j \rangle=T\chi_m$, where $\chi_m$ is the magnetic susceptibility per site, and (2) that $T\chi_m \lesssim 1$ for spin glasses.
Equation (\ref{aa1}) clearly indicates that thermodynamic limits can be obtained from Monte Carlo calculations for systems of various sizes in which contributions from the outer space are disregarded. Finally, explicit numerical evidence, Fig. \ref{ff15}, to this effect is also given.

\begin{figure}[!t]
\includegraphics*[width=75mm]{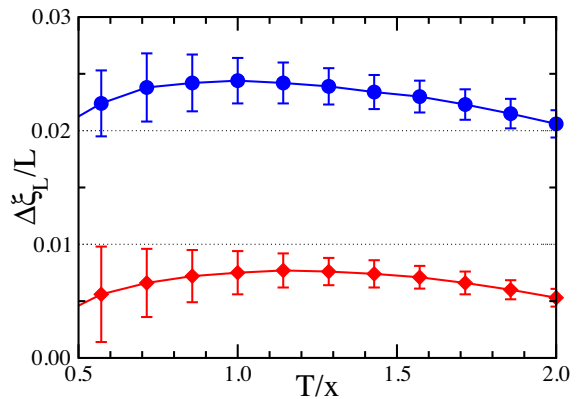}
\caption{(Color online) Semilog plots of  $\Delta \xi_L /L$ vs $T/x$, where the $\Delta \xi_L /L$ is the difference between correlation lengths we report in this paper and correlation lengths that obtain when Ewald sums are included  for $x=0.35$, ($\blacklozenge$) $L=4$ and ($\bullet$) $L=8$. These data points follow from averages over $10^4$ and $5\times 10^3$ systems samples, for $L=4$ and $L=6$, respectively. The same sample realizations were used for the calculations with and without Ewald sums.}
\label{ff15}
\end{figure}

To begin, let $h=\sum_jT_{ij} s_j$ ($\Delta h=\sum_jT_{ij} s_j$) be the sum is over all occupied sites within (outside) a cubic box of $L\times L\times L$ sites, centered on $i$.
Therefore,
\begin{equation}
 {\Delta h^2} =\sum_{n,m} T_{in}T_{im}   s_n    s_m 
\label{dh2}
\end{equation}
where the double sum is over all occupied sites in the outer space. Let
\begin{equation}
f(\bf{r}_n)=\sum_j\frac{\varepsilon_aa^3}{\mid \bf{r}_n+\bf{R}_j\mid ^3} \left[ 1-\frac{3(z_n+Z_j)^2}{\mid \bf{r}_n+\bf{R}_j\mid ^2} \right],
\end{equation}
where $ \bf{R}_j$ is the position of the outer $jth$ box, $\bf{r}_n$ is the $n$-th site's position with respect to the center of the box, and the sum is over all outer boxes. Equation (\ref{dh2}) then becomes,
\begin{equation}
{\Delta h^2}  =\sum_{n,m} f({\bf r}_n)f({\bf r}_m)  s_n   s_m  .
\end{equation}
where the sum is over all occupied sites within our system of interest. We now replace $s_n$ by $\langle s_n \rangle +\delta s_n$, and similarly for $s_m$, in the equation above. Now, it can be checked straightforwardly (i) that $\sum_m f({\bf r}_m) \langle s_n \rangle =0$ if  $\langle s_n \rangle$ is either independent of $n$ (which would not hold for a ferromagnet with domains) and (ii) that $\sum_m f({\bf r}_m) \langle s_n \rangle \rightarrow 0$ as $L\rightarrow \infty$ if $\langle s_n \rangle$ follows an antiferromagnetic order (which, for up and down spins with dipolar interactions on a SC lattice, is a checkerboard-like arrangement of up and down ferromagnetic columns). Performing thermal and disorder averages over the above equation, one then obtains, 
\begin{equation}
\langle {\Delta h^2} \rangle \rightarrow \sum_{n,m} f({\bf r}_n)f({\bf r}_m) \langle   \delta s_n    \delta s_m \rangle .
\end{equation}
as $L\rightarrow \infty$. Now, $f(\bf{r})$ varies smoothly within the system, whence
\begin{equation}
\langle {\Delta h^2} \rangle \rightarrow  \sum_{n} [ f({\bf r}_n) ]^2 \sum_m\langle   \delta s_n    \delta s_m \rangle
\end{equation}
if $\langle \delta s_n \delta s_m \rangle \simeq 0$ unless $\mid {\bf{r}}_n-{\bf{r}}_m \mid \ll L$. Finally, $\sum_{n} [ f({\bf r}_n) ]^2 = x b \varepsilon_a^2/L^3$, where $b\simeq 7.6$ if $L\gg1$, as follows straightforwardly by numerical integration.
Replacement of $\sum_m\langle  \delta s_n   \delta s_m \rangle $ by $T\chi_m$ gives  Eq.  (\ref{aa1}) if $T\chi_m$ is finite.  For all the parameters used in our MC calculations, we have found that $T \chi_{m} \lesssim 1$.}

The difference $\Delta \xi_L/L$ between the correlation lengths we report and the ones obtained when Ewald sums 
\cite{ewald} are included, for two system sizes, are exhibited in Fig. \ref{ff15}. The same sample realizations were used for the calculations with and without Ewald sums. This explains why we can show in Fig. \ref{ff15} values for $\Delta \xi_L/L$ that are smaller than the statistical errors  
given for $\xi_L/L$ (see Fig. \ref{10xivsN}) for $L=6$. 
The results are clearly consistent with a $\Delta \xi_L/L$ that vanishes in the thermodynamic limit.

\end{document}